\documentstyle[12pt]{article}

\def\II{{\rm 1\!\!I}}

\def\sinh{{\rm sinh\,}}

\def\tanh{{\rm tanh\,}}
\def\log{{\rm log\,}}

\def\min{{\rm min\,}}

\def\be{\begin{equation}}
\def\ee{\end{equation}}
\def\bea{\begin{eqnarray}}
\def\eea{\end{eqnarray}}
\newtheorem{theorem}{Theorem}

\newtheorem{corollary}{Corollary}

\begin{document}

\begin{titlepage}
\null\vskip-3truecm\hspace*{6.5truecm}{\hrulefill}\par\vskip-4truemm\par
\hspace*{6.5truecm}{\hrulefill}\par\vskip5mm\par
\hspace*{6.5truecm}{{\large\sf University of Greifswald (Oct., 1997)}}
\vskip4mm\par\hspace*{6.5truecm}{\hrulefill}\par\vskip-4truemm\par
\hspace*{6.5truecm}{\hrulefill}\par
\bigskip\bigskip\hspace*{6.5truecm}
\par
\hspace*{6.5truecm}solv-int/9710009
\bigskip\par
\hspace*{6.5truecm}{\sf submitted to:}
{\sf J. Mathematical Physics}\par
\vfill
\centerline{\LARGE\bf A new explicit expression}
\medskip
\centerline{\LARGE\bf for the Korteweg-De Vries hierarchy}
\bigskip
\bigskip
\centerline{\large\bf Ivan G. Avramidi
\footnote{On leave of absence from Research Institute for Physics, 
Rostov State University,  Stachki 194, 344104 Rostov-on-Don, Russia.
E-mail: avramidi@rz.uni-greifswald.de}
{\rm and} Rainer Schimming
\footnote{\rm E-mail: schimmin@rz.uni-greifswald.de}}
\bigskip
\centerline{\it Institut f\"ur Mathematik und Informatik, Ernst-Moritz-Arndt Universit\"at Greifswald}
\centerline{\it D--17489 Greifswald, Germany}
\bigskip
\centerline{October  14, 1997}
\vfill

{\narrower
\par
We derive an improved fully explicit expression for the 
right-hand sides of the 
matrix KdV hierarchy using the relation to the heat 
kernel of the one-dimensional Schr\"odinger operator.
Our method of "matrix elements" produces, moreover, 
an explicit expression for the powers of a Schr\"odinger-like differential operator of any order.
\par}
\vfill

\end{titlepage}

\section{Introduction}

The {\it Korteweg-de Vries} (abbreviated KdV henceforth)
equation
\be
{\partial u\over\partial t}={\partial^3 u\over \partial x^3}
+6u{\partial u\over \partial x}
\ee
for a function $u=u(x,t)$ of position $x$ and time $t$ naturally extends to an infinite sequence of partial differential equations of solitonic character
\cite{lax68}
\be
{\partial u\over\partial t}={\partial\over \partial x}G_k[u]
\qquad (k=1,2,\dots),
\label{1c}
\ee
where each $G_k=G_k[u]$ is some differential polynomial in $u$ with respect to $x$,
i.e. a polynomial in $u$ and its derivatives
\be
u_n:={\partial^n u\over \partial x^n},\qquad 
(n=1,2,\dots).
\ee
The sequence begins with
\bea
G_1&=&u,
\nonumber\\
G_2&=&u_2+3u^2,
\nonumber\\
G_3&=&u_4+10 uu_2+5u_1^2+10u^3, \quad \dots
\eea
While usually the {\it KdV hierarchy} is defined through a 
{\it recursive} rule for the
$G_k$, we derived in our papers \cite{schim95,jmp95b}
an {\it explicit} expression for the latter. The main aim of the 
present paper is to improve this expression by decomposing it in homogeneous parts.
Thus, the rest of impliciteness in \cite{schim95,jmp95b}
is overcome. We derive the fully explicit expression by means 
of a "{\it method of matrix elements}", developed in \cite{npb91,phd86},
which will be described shortly in section 2.
The same method produces an expression for the powers
$L^k$ of the {\it one-dimensional Schr\"odinger operator}
\be
L={\partial^2\over\partial x^2}+u
\label{2c}
\ee
and gives some knowledge on the differential operators $A_k$ in the 
{\it Lax representation}
\be
{\partial L\over \partial t}=[A_k,L],
\qquad (k=1,2,\dots)
\ee
of the KdV hierarchy \cite{lax68}. More generally, the method gives, as a side-result,
an expression for the powers $L^k$ of an 
{\it $r$-th order Schr\"odinger-like operator}
\be
L_r={\partial^r\over\partial x^r}+u.
\ee
A particularly convenient formula, in terms of the so called 
{\it Bell differential polynomials}, is valid for $r=1$.

While the paper \cite{schim95} treated a {\it scalar} potential $u=u(x,t)$
only, the derivation in \cite{jmp95b} and in the present paper holds for the {\it matrix case} too, where $u=u(x,t)$ is
a $N\times N$ matrix valued function.
Notice that the matrix KdV hierarchy (\ref{1}) has solitonic character if $u$ is
a Hermitian (in particular, a real symmetric) matrix:
the papers \cite{wadati74,calogero77,olmedilla81,olmedilla85,alonso82} showed that the inverse scattering method for
solving Cauchy's problem works well then. 

It is well known (cf., e.g., \cite{npb91,jmp95b,hurt83,qftext96})
that the {\it heat kernel} $K=K(s;x,x')$, i.e. 
the fundamental solution of the heat equation
\be
\left({\partial \over\partial s}-L\right)K=0,
\ee
to the one dimensional Schr\"odinger operator (\ref{2c})
has an asymptotic expansion as $t\to 0$ of the form:
\be
K=K(s;x,x')\sim(4\pi s)^{-1/2}\exp\left(-{(x-x')^2\over 4s}\right)
\sum\limits_{k=0}^\infty{1\over k!}a_k(x,x')s^k.
\ee
The coefficients $a_k=a_k(x,x')$ satisfy the recursive differential equations
\be
\left(1+{1\over k}D\right)a_k=L a_{k-1},
\qquad (k=1,2,\dots),
\qquad
a_0:=1,
\label{3c}
\ee
where we abbreviate
\be
D:=(x-x'){\partial\over\partial x}.
\ee
It is further known \cite{mckean75,perelomov76,jmp95b} that the right-hand sides of the KdV hierarchy are 
determined by the diagonal values of the heat kernel coefficients:
\be
G_k[u](x)={1\over 2}{2k \choose k}a_k(x,x).
\label{14a}
\ee

We will evaluate the recursion system (\ref{3c})
in order to calculate the diagonal values $a_k(x,x)$. In fact, we do more and calculate
all Taylor coefficients
\be
<n|a_k>:={\partial^n\over\partial x^n} a_k(x,x')\big|_{x=x'}.
\ee


\section{The method of matrix elements}

Let us briefly describe a "{\it method of matrix elements}" 
as developed in \cite{phd86,npb91,qftext96}.
Let us fix a point $x'$ on the real line ${\bf R}$ and consider
a sufficiently small interval around $x'$, say ${\bf I}=[x'-R,x'+R]$,
where $R$ is a sufficiently small positive constant. 
We consider the space $C^\omega(I)$ of analytic functions $f=f(x,x')$
of the real variable $x\in {\bf I}$ and their Taylor series
\be
f(x,x')=\sum\limits_{n=0}^\infty {1\over n!}(x-x')^n
{\partial^n \over \partial x^n}f(x,x')\Bigg\vert_{x=x'}.
\label{a1}
\ee
We rewrite the Taylor formula (\ref{a1}) in the form
 \be
f\equiv |f>=\sum\limits_{n=0}^\infty |n><n|f>,
\label{a1a}
\ee
where
\be
|n>:={1\over n!}(x-x')^n, \qquad (n=0,1,2,\dots),
\label{a5}
\ee
\be
<n|f>:={\partial^n \over \partial x^n}f(x,x')\Bigg\vert_{x=x'}.
\ee
In a sense, the functions $|n>$ form a basis of the space $C^\omega({\bf I})$
and the $<n|$ are the corresponding {\it dual} linear 
functionals $<n|:\ C^\omega({\bf I})
\to {\bf R}$. Taylor's theorem can formally be expressed as
a {\it completeness relation}
\be
\sum_{n=0}^\infty |n><n|=\II,
\label{a7}
\ee
where $\II$ denotes the identical operator.
The duality of the systems $<m|$ and $|n>$ is expressed by
\be
<m|n>=\delta_{mn},
\ee
where $\delta_{mn}$ is the Kronecker symbol.

Let us consider a linear differential operator of order $r$ with respect to 
$x\in {\bf I}$
$$
L=\sum_{j=1}^r u_j(x,x'){\partial^j\over\partial x^j}
$$
with analytic
coefficients $u_j=u_j(x,x')$.
Such an operator can be characterized by
its {\it matrix elements}
\be
<m|L|n>:=<m|(L|n>)
=\sum_{j=1}^{\min{(r,n)}}
{m\choose n-j}<m-n+j|u_j>
\ee
Namely, there holds
\be
L=\sum_{n,m\ge 0}^\infty |m><m|L|n><n|,
\label{a6}
\ee
which is just an abbreviation of
\be
L|f>=\sum_{n,m\ge 0}^\infty |m><m|L|n><n|f>.
\ee
Note that for differential operators of order $r$ the matrix elements $<m|L|n>$ vanish for $n\ge m+r+1$:
\be
<m|L|n>=0 \qquad {\rm for}\ n\ge m+r+1.
\label{a8}
\ee
Therefore, the sum over $n$ in (\ref{a6}) is always finite! The infinite sum over $m$ 
converges because of the analyticity of the coefficients $u_j$ of the operator $L$.
This is a special property of differential operators and it is not 
necessarily valid for
pseudodifferential operators.
 
Formally, (\ref{a6}) is obtained by application of  the identity operator 
(\ref{a7}) on both sides.
By inserting the completeness relation (\ref{a7})  one can show that the matrix elements of the product of two operators $L_1L_2$ result from the matrix multiplication of the matrix elements of $L_1$ and $L_2$:
\be
<m|L_2L_1|n>=\sum_{l=0}^\infty <m|L_2|l><l|L_1|n>.
\ee
Again, from the property (\ref{a8}), we see that this sum is finite and there is no
problem of convergence. 
The same argument applies to the product $L_k\cdots L_2L_1$ of operators 
$L_1, L_2, \cdots, L_k$:
\bea
<m|L_k\cdots L_2L_1|n>&=&\sum_{n_1,n_2,\dots,n_{k-1}\ge 0}<m|L_k|n_{k-1}>
<n_{k-1}|L_{k-1}|n_{k-2}>\cdots
\nonumber\\
&\times&\cdots
<n_2|L_2|n_{1}><n_1|L_1|n>.
\eea

Let us note that the method of matrix elements applies also to the more general
case of smooth functions, i.e. to the space $C^\infty({\bf I})$.
In this case the Taylor series does not, in general, converge
but is an {\it asymptotic} expansion. Thus, all the equations of this section hold
also for operators with smooth , not necessarily analytic, coefficients,
if one replaces the sign $=$ by the sign for asymptotic equality
$\sim$  in the Taylor expansion.
The equations between Taylor coefficients remain exactly valid even in this case.

Let us consider now some examples of linear operators.
\begin{enumerate}
\item
We already know that the identical operator $\II$ has the matrix elements
\be
<m|\II|n>=<m|n>=\delta_{mn}
\ee

\item
The special first-order operator
\be
D:=(x-x'){\partial\over\partial x}
\ee
has the Taylor basis (\ref{a5}) as a complete system of eigenfunctions:
\be
D|n>=n|n> \qquad {\rm for}\ n=0,1,2,\dots .
\ee
As a conclusion, the matrix elements of $D$ read
\be
<m|D|n>=n\delta_{mn};
\ee
hence,
\be
D=\sum_{n=0}^\infty n|n><n|.
\ee
More generally, for an arbitrary {\it polynomial} $P$ we have
\be
P(D)=\sum_{n=0}^\infty P(n)|n><n|.
\ee
This operator can be inverted provided the polynomial does not have roots at the integer positive points, i.e. if 
$P(n)\ne 0$ {\rm for}\ $ n=0,1,2,\dots$
then 
\be
P(D)^{-1}=\sum_{n=0}^\infty {1\over P(n)}|n><n|
\label{38d}
\ee
The polynomial values $P(n)$ in the denominator even improve the convergence
if $P(D)^{-1}$ is applied to an analytic function.

\item
The {\it one-dimensional Schr\"odinger operator}
\be
L:={d^2\over dx^2}+u, \qquad u=u(x),
\label{39c}
\ee
has the matrix elements
\bea
&&<m|L|n>={m\choose n}u_{m-n} \qquad {\rm for}\ n\le m,
\label{35aa}\\
&&<m|L|m+1>=0,\\
&&<m|L|m+2>=1,\\
&&<m|L|n>=0 \qquad {\rm for}\ n\ge m+3,
\label{39aa}
\eea
where
\be
u_{n}:={d^n u\over dx^n}.
\ee

The result can also be compactly expressed by:
\be
<m|L|n>={m\choose n}u_{m-n}+<m+2|n>,
\ee
if we set
\be
{m\choose n}:=0, \qquad u_{n}:=0 \qquad {\rm for}\ n<0.
\label{a10}
\ee
It is also helpful to arrange the matrix elements to
an infinite matrix
\be
{\bf L}\equiv(<m|L|n>)=
\left(
\begin{array}{cccccccc}
u_0         & 0            &1        &0                              
&0&0&0       &\cdots\\
u_{1}&u_0      &0                &1                              
&0&0&0      &\cdots\\
u_{2}&{2\choose 1}u_{1}&u_0       &0                             
&1&0&0      &\cdots\\
u_{3}&{3\choose 1}u_{2}&{3\choose 2}u_{1}&u_0                              
&0&1&0     &\cdots\\
u_{4}&{4\choose 1}u_{3}&{4\choose 2}u_{2}&{4\choose 3}u_{1} 
&u_0&0&1      &\cdots\\
u_{5}&{5\choose 1}u_{4}&{5\choose 2}u_{3}&{5\choose 3}u_{2} 
&{5\choose 4}u_{1} &u_0&0     &\cdots\\
u_{6}&{6\choose 1}u_{5}&{6\choose 2}u_{4}&{6\choose 3}u_{3} 
&{6\choose 4}u_{2} &{6\choose 5}u_{1} &u_0    &\cdots\\
\vdots                      &\vdots                       &\vdots                      
&\vdots  &\vdots&\vdots&\vdots&\ddots
\end{array}
\right)
\label{48c}
\ee
This matrix can be expanded as an infinite sum of constant 
matrices with shifted diagonals multiplied by derivatives of $u$:
\bea
{\bf L}&=&I_{-2}+u_0I_0+u_1I_1+\cdots
\nonumber\\
&=&
\sum\limits_{j\ge 0} u_{j-2}I_{j-2}
\label{49c}
\eea
where we formally set
\be
u_{-2}:=1, \qquad u_{-1}:=0,
\label{55c}
\ee
and the proper meaning of each matrix $I_j$ is just defined by (\ref{48c}).

\item
More generally, let us consider a linear differential operator of the form
\be
L_r={\partial^r\over\partial x^r}+u, \qquad u=u(x).
\label{51c}
\ee
where $r$ is a positive integer.
We will call it 
a higher-order {\it Schr\"odinger-like operator}.
The one-dimensional Schr\"odinger operator is the special case $r=2$. 
The calculation of the matrix elements in the preceding item can be 
extended to general $r$.
We obtain
\be
<m|L_r|n>={m\choose n}u_{m-n}+<m+r|n>,
\ee
if we adopt the convention (\ref{a10}).
This defines the matrix representation of ${\bf L}_r$
\be
{\bf L}_r=<m|L_r|n>=\sum\limits_{j\ge 0} u_{j-r}I_{j-r}
\label{52c}
\ee
where now we alternatively set
\be
u_{-r}=1, \qquad u_{-r-1}=\cdots=u_{-1}=0,
\label{50c}
\ee
and the matrix $I_{-r}$ is defined by
\be
<m|I_{-r}|n>=<m+r|n>.
\ee

\item
Finally, let us consider the {\it powers} $L_r^k$ of the 
Schr\"odinger-like operator
(\ref{51c}).
Clearly, the matrix elements $<m|L_r^k|n>$ of the powers of the
operator $L_r$ are given
just by the $k$-th power of the infinite matrix ${\bf L}_r$ (\ref{52c}).
Thus
\be
L^k_r=\sum\limits_{m,n\ge 0}|m><m|L^k_r|n><n|,
\ee
where
\be
<m|L^k_r|n>=\sum\limits_{j_1,\dots,j_k\ge 0}
h({\bf j})u_{j_1-r}\dots u_{j_k-r}.
\label{50hhh}
\ee
Here ${\bf j}=(j_1,\dots,j_k)$ is a multiindex 
formed from non-negative integers $j_1$, $j_2$, $\dots$, $j_k$
and 
the constants $h({\bf j})$ are determined by the products
of the matrices $I_j$:
\be
h({\bf j})=
<m|I_{j_1-r}\cdots I_{j_k-r}|n>
\ee

\end{enumerate}

\section{Derivation of the KdV polynomials}

Let us apply a special multi-index formalism in order to present our result:
we set
\be
{\bf m}:=(m_1,m_2,\dots,m_d),
\ee
\be
|{\bf m}|_{p}:=m_1+m_2+\cdots+m_p, \qquad {\rm for}\ p=1,2,\dots,d ,
\ee
\be
|{\bf m}|:=|{\bf m}|_{d}=m_1+m_2+\cdots+m_d,
\ee

\begin{theorem}
The Taylor coefficients of the heat kernel coefficients 
$a_k=a_k(x,x')$ to the one-dimensional Schr\"odinger operator
(\ref{2c}) read
\be
<n|a_k>=\sum_{d=1}^k\sum_{|{\bf m}|=n+2k-2d}
c({\bf m})u_{m_d}\cdots u_{m_2}u_{m_1}
\label{1}
\ee
where the innner sum runs over integers $m_1\ge 0, m_2\ge 0, \dots, m_d\ge 0$ such that 
\be
|{\bf m}|\equiv m_1+m_2+\cdots+m_d=n+2k-2d
\ee
and where the numerical coefficients read
\bea
c({\bf m})&\equiv& c(m_1,m_2,\dots,m_d)
\nonumber\\[12pt]
&=&
\sum_{i_1,\cdots,i_{d-1}}\prod_{p=1}^d {i_p\choose i_{p-1}}
{{|{\bf m}|_p-2i_{p-1}+2p\choose m_p}}
{{|{\bf m}|_p-i_{p-1}+2p+1\choose i_p-i_{p-1}}}^{-1}.
\label{15}
\eea
The sum in (\ref{15}) runs over integers $i_1, i_2, \dots, i_{d-1}$ such that
\be
0\equiv i_0<i_1<i_2<\cdots<i_{d-1}<i_d\equiv k,
\ee
\be
2i_p\le |{\bf m}|_{p+1}+2p.
\ee

\end{theorem}

\paragraph{Proof.}
The recursive differential equations (\ref{3c}) have the symbolic
solution
\be
a_k=\left(1+{1\over k}D\right)^{-1}L\cdots \left(1+{1\over 2}D\right)^{-1}
L\left(1+{1\over 1}D\right)^{-1}L\cdot 1
\ee
This can be translated, by use of the eq. (\ref{38d}),
into a formula in terms of matrix elements:
\bea
<n|a_k>=\sum_{n_1,\dots,n_{k-1}}
\left(1+{n_k\over k}\right)^{-1}\cdots \left(1+{n_2\over 2}\right)^{-1}
\left(1+{n_1\over 1}\right)^{-1}
\nonumber\\
\times<n_k|L|n_{k-1}>\cdots<n_2|L|n_1><n_1|L|0>,
\label{2a}
\eea
where $n_k\equiv n$ and the summation indices $n_1,n_2,\dots,n_{k-1}$ run through $0,1,2,\dots$.

Recall the values (\ref{35aa})-(\ref{39aa}) of the matrix elements of $L$.
Although the sum in (\ref{2a}) is formally an infinite one, it is effectively finite. 
Let us call a matrix element $<m|L|n>$ {\it essential} if $n\le m$ and
{\it non-essential} if $n\ge m+1$. The number of essential elements in a term
of (\ref{2a}) equals the {\it degree} $d$ of the differential monomial in $u$.
Let us analyze a typical term of degree $d=N+1$ in (\ref{2a}).
Clearly, $<n_1|L|0>$ is essential since $n_1\ge 0$. Let 
\be
<n_{i_N+1}|L|n_{i_N}>, \ \dots, \ <n_{i_1+1}|L|n_{1}>, \ <n_{i_1}|L|0>.
\ee
denote all the essential elements, numerated from the right to the left. 
The space between two essential elements is filled by a chain of unessential 
matrix elements, which can be written as Kronecker symbols. 
The unessential chain between $<n_{i_1+1}|L|n_{i_1}>$
and \hbox{$<n_1|L|0>$} reads
\be
<n_{i_1}+2|n_{i_1}>\cdots<n_3+2|n_2><n_2+2|n_1>.
\ee
This product of Kronecker symbols cancels the summations over $n_2,n_3,\dots,n_{i_1}$ by
fixing these running indices:
\bea
n_2&=&n_1-2,
\nonumber\\
n_3&=&n_2-2=n_1-4,
\nonumber\\
&\ldots&
\nonumber\\
n_{i_1}&=&n_{i_1-1}-2=\dots=n_1-2(i_1-1).
\label{3}
\eea
The chain under consideration produces the numerical coefficient
\bea
c_0&:=&\left(1+{n_{i_1}\over i_1}\right)^{-1}\cdots 
\left(1+{n_{3}\over 3}\right)^{-1}\left(1+{n_{2}\over 2}\right)^{-1}
\left(1+{n_{1}\over 1}\right)^{-1}
\nonumber\\
&\equiv&{i_1\over n_{i_1}+i_1}\cdots 
{3\over n_{3}+3}\cdot{2\over n_{2}+2}\cdot{1\over n_{1}+1}.
\eea
By virtue of (\ref{3}), it becomes
\be
c_0={{{n_{1}+1}\choose i_1}}^{-1}.
\ee
The next unessential chain between $<n_{i_2+1}|L|n_{i_2}>$
and $<n_{i_1+1}|L|n_{i_1}>$ reads
\be
<n_{i_2}+2|n_{i_2-1}>\cdots<n_{i_1+3}+2|n_{i_1+2}><n_{i_1+2}+2|n_{i_1+1}>
\ee
It cancels the summation over $n_{i_1+2}, n_{i_1+3}, \dots, n_{i_2}$ by
fixing
\bea
n_{i_1+2}&=&n_{i_1+1}-2,
\nonumber\\
n_{i_1+3}&=&n_{i_1+2}-2=n_{i_1+1}-4,
\nonumber\\
&\ldots&\nonumber\\
n_{i_2}&=&n_{i_2-1}-2=\dots=n_{i_1+1}-2(i_2-i_1-1)
\label{4}
\eea
and produces the numerical coefficient
\bea
c_1&:=&\left(1+{n_{i_2}\over i_2}\right)^{-1}\cdots 
\left(1+{n_{i_1+3}\over i_1+3}\right)^{-1}
\left(1+{n_{i_1+2}\over i_1+2}\right)^{-1}
\left(1+{n_{i_1+1}\over i_1+1}\right)^{-1}
\nonumber\\
&\equiv&{i_2\over n_{i_2}+i_2}\cdots 
{3\over n_{i_1+3}+i_1+3}
\cdot{i_1+2\over n_{i_1+2}+i_1+2}\cdot{i_1+1\over n_{i_1+1}+i_1+1}
\eea
By virtue of (\ref{4}), the latter becomes
\be
c_1={{i_2\choose i_1}}{{{n_{i_1+1}+i_1+1}\choose i_2-i_1}}^{-1}.
\ee
The procedure carries on this way; the last unessential chain fixes
\bea
n_{i_N+2}&=&n_{i_N+1}-2,
\nonumber\\
n_{i_N+3}&=&n_{i_N+2}-2=n_{i_N+1}-4,
\nonumber\\
&\ldots&\nonumber\\
n_{k-1}&=&n_{k-2}-2=\dots=n_{i_N+1}-2(k-i_N-2)
\nonumber\\
n_{k}\equiv n&=&\dots=n_{i_N+1}-2(k-i_N-1),
\label{5}
\eea
and produces the numerical coefficient
\be
c_N={{k\choose i_N}}{{n_{i_N+1}+i_N+1\choose k-i_N}}^{-1}.
\ee
After the reduction procedure the typical term in ({\ref{2a}}) reads
\be
c_N\cdots c_1c_0
<n_{i_N+1}|L|n_{i_N}>\cdots<n_{i_1+1}|L|n_{1}><n_{1}|L|0>
\ee
and the summation indices are now
$
N,\ i_1,\ i_2,\ \dots,\ i_N, \ n_1, \ n_{i_1+1}, \ n_{i_2+1},\dots,\  n_{i_{N-1}+1}.
$
Notice, further, that $n_k=n$ is fixed. 

In order to get a better expression for the essential matrix elements 
it is convenient to pass to new summation indices
$
N,\ i_1,\ i_2,\ \dots,\ i_N, \ m_1, \ m_2, \ \dots,\ m_{N+1},
$
where
\bea
m_1&:=&n_1,
\nonumber\\
m_2&:=&n_{i_1+1}-n_{i_1}=n_{i_{1}+1}-n_1+2(i_1-1),
\nonumber\\
m_{3}&:=&n_{i_2+1}-n_{i_2}=n_{i_2+1}-n_{i_1+1}+2(i_2-i_1-1),
\nonumber\\
&\ldots&\nonumber\\
m_{N+1}&:=&n_{i_N+1}-n_{i_N}=n_{i_N+1}-n_{i_{N-1}+1}
+2(i_N-i_{N-1}-1).
\label{7}
\eea
The old summation indices are expressed in terms of the new ones by
\bea
n_1&=&m_1,
\nonumber\\
n_{i_1+1}&=&m_1+m_2-2i_1+2,
\nonumber\\
n_{i_2+1}&=&m_1+m_2+m_3-2i_2+4,
\nonumber\\
&\ldots&\nonumber\\
n_{i_N+1}&=&m_1+m_2+\cdots+m_{N+1}-2i_N+2N.
\label{8}
\eea
Hence
\bea
c_0&=&{{m_1+1\choose i_1}}^{-1},
\nonumber\\
c_1&=&{{i_2\choose i_1}}{{m_1+m_2-i_3+3\choose i_2-i_1}}^{-1},
\nonumber\\
&\ldots&
\nonumber\\
c_N&=&{{k\choose i_N}}{{m_1+\cdots+m_{N+1}-i_N+2N+1\choose k-i_N}}^{-1}.
\eea
Some restriction for $m_1, m_2,\dots, m_{N+1}$ is to be considered:
\be
m_1+m_2+\cdots+m_{N+1}=n+2(k-N-1);
\ee
it follows from (\ref{7}).
Besides, all $m_i$ are non-negative and we have to require
\be
0\le 2i_p\le m_1+m_2+\cdots+m_{p+1}+2p.
\ee
Finally, we express the essential matrix elements in terms of $m_1$, $m_2$, $\dots$, $m_{N+1}$:
\bea
<n_1|L|0>&=&{m_1\choose 0}u_{m_1},
\nonumber\\
<n_{i_1+1}|L|n_{i_1}>&=&{m_1+m_2-2i_1+2\choose m_2}u_{m_2},
\nonumber\\
<n_{i_2+1}|L|n_{i_2}>&=&{m_1+m_2+m_3-2i_2+4\choose m_3}u_{m_3},
\nonumber\\
&\ldots&\\
<n_{i_N+1}|L|n_{i_N}>&=&{m_1+\cdots+m_{N+1}-2i_N+2N\choose m_{N+1}}u_{m_{N+1}}.
\eea
We complete the proof by setting
$d:=N+1, i_0:=0, i_d:=k$
and summation over $d=1,2,\dots,k$.
The asserted formula (\ref{1})--(\ref{15}) follows.

The outer summation in 
(\ref{1}) is a decomposition into homogeneous parts.
Note, further, that the differential polynomial $<n|a_k>$ in $u$ is {\it isobaric}, 
that means every term has the same {\it weight}
\be
|{\bf m}|\equiv m_1+m_2+\cdots+m_d=n+2(k-d).
\ee

Theorem 1 is a very special one-dimensional case of the general
formula for the Taylor coefficients of the heat kernel coefficients to an arbitrary
second-order partial differential operator of Laplace type obtained by one of the  authors (I.G.A.) 
in \cite{phd86,npb91}. 
We present here for the first time the complete proof; in \cite{phd86,npb91} 
the combinatorial details were omitted. 

Specializing to
\be
<0|a_k>=a_k(x,x),
\ee
and considering (\ref{14a}) we obtain

\begin{theorem}
The right-hand sides of the KdV hierarchy (\ref{1c}) are given by
\be
G_k[u]={1\over 2}{2k\choose k}
\sum_{d=1}^k\sum_{|{\bf m}|=2k-2d}
c({\bf m})u_{m_d}\cdots u_{m_2}u_{m_1}
\label{1d}
\ee
where the multiindex ${\bf m}$ and the numerical coefficients $c({\bf m})$
have the same meaning as in Theorem 1.

\end{theorem}

Notice that (\ref{1d}) is valid for the {\it matrix KdV hierarchy}, since 
our proof did not make use of a commutative law.
The convential {\it scalar KdV hierarchy}, where commutativity 
is assumed, emerges as a special case.

\section{On the structure of the KdV polynomials}

The explicit expression (\ref{1d}) is well suited to be evaluated by means of computer
algebra. It is not practical to calculate higher terms by hand. 
Also, (\ref{1d}) conceals that some terms in $G_k[u]$ have a relatively simple structure, as we are going to discuss now.
The asymptotic expansion of the heat kernel diagonal
\be
K(s;x,x)\sim (4\pi s)^{-1/2}\sum\limits_{k\ge 0}{1\over k!}a_k(x,x)\,s^k
 =(4\pi s)^{-1/2}2\sum\limits_{k\ge 0}{k!\over (2k)!} G_k[u]\,s^k
\label{96j}
\ee
is a generating function for the sequence of the KdV polynomials $G_k$.
It is well known, that the diagonal value $K(s;x,x)$ of the of the heat kernel
obeys
\be
K(s;x,x)=e^{su_0}\tilde K(s;x,x),
\ee
where $\tilde K$ is the heat kernel diagonal calculated for $u_o=0$.
As a consequence, there holds
\be
a_k=\sum_{i=0}^k {k\choose i}u_0^{i}\tilde a_{k-i},
\ee
and
\be
G_k=\sum_{i=0}^k {k-{1\over 2}\choose i}(4u_0)^{i}\tilde G_{k-i},
\ee
where
\be
\tilde a_k=a_k\big|_{u_0=0},\qquad
\tilde G_k=G_k\big|_{u_0=0}.
\ee
Notice that
\be
{k-{1\over 2}\choose i}={\Gamma\left(k+{1\over 2}\right)\over 
i!\Gamma\left(k-i+{1\over 2}\right)}
={1\over 2^{2i}}{2k\choose 2i}{2i\choose i}{k\choose i}^{-1}.
\ee
Herefrom we find the terms of highest degree and lowest order in $G_k$ as
far as we want:
\be
60{2k\choose k}^{-1}G_k=30 u_0^k+10{k\choose 2}u_0^{k-2}u_2+
3{k\choose 3}u_0^{k-3}(u_4+5u_1^2)+\cdots
\ee

The terms of lowest degree and highest order are found by means of some other general relation.
Namely, we  have shown in \cite{jmp95b} that for the matrix KdV hierarchy the recursion
\be
G_k=2\sum_{i=1}^k Z_{2i-1}G_{k-i}, \qquad (k=1,2,\dots),
\ee
holds, where the differential polynomials $Z_k=Z_k[u]$ on their side are
recursively defined by
\be
Z_{k+1}:={d\over dx} Z_k+\sum_{i=1}^{k-1}Z_iZ_{k-i},
\ee
\be
Z_1:=u_0,
\ee
(The proof is far from being trivial in the non-commutative case.)
This enabled us in \cite{jmp95b} to find
\be
G_k=u_{2k-2}+\sum_{i=0}^{2k-4}\left\{{2k-2\choose i+1}+(-1)^i
\right\}u_iu_{2k-4-i}+\cdots,
\ee
where the points indicate terms of higher degree.

Let us add here another result on the structure of the KdV hierarchy: we find,
in the commutative case, all monomials in $G_k$ which are built solely from
$u_0$, $u_1$ and $u_2$.

The proof uses again diagonal values of the heat kernel as generating function and the
fact that $u_0$, $u_1$ $u_2$ commute with each other. 
Besides, we will use the Bernoulli numbers $B_n$, $(n=0,1,2,\dots)$,
(cf., e.g., \cite{erdelyi53}) defined recursively by
\be
\sum_{k=0}^n{n+1\choose k}B_k=0, \qquad B_0:=1.
\ee
The first Bernoulli numbers read: $B_1=-{1\over 2}$, $B_2=-{1\over 6}$,
$B_3=0$, $B_4=-{1\over 30}$. Generally, $B_{2k+1}=0$, 
$B_{2k}=(-1)^{k+1}|B_{2k}|\ne 0$ for $k\ge 1$.

\begin{theorem}
For the scalar KdV hierarchy there holds
\be
G_k[u]\equiv{(2k)!\over 2k!k!}a_k(x,x)=
\sum\limits_{N=1}^k{(2k)!\over 2k!N!}
\sum\limits_{|{\bf m}|=k}{b_{m_1}\over m_1!}\cdots {b_{m_N}\over m_N!}
+\cdots
\label{96hhh}
\ee
where the inner sum runs over integer $m_1\ge 1,\dots, m_N\ge 1$
such that $|{\bf m}|\equiv m_1+\cdots+m_N=k$,
the monomials $b_k$ are defined by
\bea
b_1&=&u_0,
\label{96hh}\\[12pt]
b_{2n}&=&{1\over n}{2^{3n-2}}|B_{2n}| u_2^n,\\[12pt]
b_{2n-1}&=&{1\over n}{2^{n-1}(2^{2n}-1)}
|B_{2n}|u_2^{n-2}u_1^2, \qquad (n\ge 2),
\label{97hhh}
\eea
and the dots indicate terms
which effectively contain higher derivatives $u_3$, $u_4$ etc. of $u$.

\end{theorem}

\paragraph{Proof.}
Actually, we have to evaluate the asymptotic expansion of the heat kernel
for a quadratic potential (harmonic oscillator)
\be
u(x)=u_0(x')+u_1(x')(x-x')+{1\over 2}u_2(x')(x-x')^2.
\ee
The heat kernel diagonal in this particular case reads
(up to exponentially small terms that do not contribute to the generating
function)
\be
K(s;x,x)\sim
(4\pi s)^{-1/2}\exp\left\{su_0+\Phi(s)+{1\over 4}s^3u_1^2\Psi(s)\right\}
\ee
where
\be
\Phi(s)=-{1\over 2}\log\left({\sinh (2s\omega)\over 2s\omega}\right),
\ee
\be
\Psi(s)={s\omega-\tanh (s\omega)\over (s\omega)^3}
\ee
and
\be
\omega\equiv \sqrt{-{u_2/2}}
\ee
Actually, this is the well-known 
heat kernel diagonal 
of a the one-dimensional harmonic oscillator,
which can be read from the literature 
(cf., e.g. \cite{hurt83,jmp95c}).
Expanding the heat kernel diagonal one finds all the coefficients $a_k(x,x)$
in the considered approximation.
The Taylor expansions of the functions $\Phi(t)$ and $\Psi(t)$ read
\be
\Phi(s)=\sum_{n\ge 1}{2^{3n-2}|B_{2n}|\over n (2n)!} u_2^n s^{2n} ,
\ee
\be
\Psi(s)=\sum_{n\ge 2}{2^{n+2}(2^{2n}-1)|B_{2n}|\over (2n)!}
u_2^{n-2}s^{2n-4},
\ee
where $B_n$ are the Bernoulli numbers.
Obviously, these series have only positive coefficients.
We obtain the heat kernel diagonal in the form
\be
K(s;x,x)\sim (4\pi s)^{-1/2}\exp\left(\sum_{n\ge 1}{s^n\over n!}b_n\right),
\ee
where the coefficients $b_n$ are defined by (\ref{96hh})-(\ref{97hhh}).
Note that all $b_n$ are positive.
The heat kernel coefficients $a_k(x,x)$ and, therefore, 
the differential polynomials $G_k$
(\ref{96j}) are determined by the expansion of the 
exponent, which completes the proof.

Let us note that the heat kernel coefficients $a_k(x,x)$, in the 
above approximation, are given by the so-called 
{\it Bell polynomials} ${\cal B}_k(y_1,\dots,y_k)$.
Namely,
the Bell polynomials are defined 
by their generating function
\be
\sum_{k=0}^\infty {s^k\over k!}{\cal B}_k
=\exp\left(\sum_{n=1}^\infty{s^n\over n!}y_n\right).
\ee
Therefore, the formula (\ref{96hhh}) can be also rewritten in 
the form
\be
G_k[u]={1\over 2}{2k\choose k}{\cal B}_k\left(b_1,\dots,b_k\right).
\ee
Several explicit expressions for the Bell polynomials are known, 
in particular,
\be
{\cal B}_n(y_1,\dots,y_d)=\sum_{d=1}^n\sum_{|{\bf m}|=n}\prod_{p=1}^d
{|{\bf m}|_p-1\choose |{\bf m}|_{p-1}}y_{m_p}
\label{112hhh}
\ee
for $n\ge 1$ \cite{schimming96}.
Here, as usual, ${\bf m}=(m_1,\dots,m_d)$ with non-negative integers 
$m_1,\dots,m_d$ and
\be
|{\bf m}|_p:=m_1+\cdots+m_p, \qquad {\rm for}\ p=1,\dots,d
\ee
\be
|{\bf m}|_0:=0,\qquad |{\bf m}|:=|{\bf m}|_d
\ee
The expression (\ref{112hhh}) is valid for generally non-commuting variables $y_1,y_2,\dots$.
Another expression, which holds only in the commutative case, is due to Faa di Bruno (cf. e.g. \cite{schimming96}):
\be
{\cal B}_n(y_1,\dots,y_n)=n!\sum_{||{\bf m}||=n}\prod_{j=1}^n {1\over m_j}\left({y_j\over j!}\right)^{m_j}
\ee
where the sum runs over all non-negative integers $m_1,\dots,m_n$ such that
\be
||{\bf m}||:=m_1+2m_2+\dots nm_n=n
\ee

One should point out here that the theorem 3. is a particular case of a general
result \cite{jmp95c} of one of the authors (I.G.A.).
In that paper the heat kernel diagonal to a Laplace type operator
$L=-g^{\mu\nu}\nabla^V_\mu\nabla^V_\nu+Q$ 
acting on sections of a vector bundle $V$ over a flat $n$-dimensional 
manifold with a metric $g$ and the bundle connection $\nabla^V$ was studied
and the part of the heat kernel diagonal that depends solely on the
curvature ${\cal R}_{\mu\nu}=[\nabla_\mu,\nabla_\nu]$ 
of the bundle connection, the potential 
$Q$ and its first $\nabla_\mu Q$ and second $\nabla_\mu\nabla_\nu Q$ 
derivatives is calculated in a closed form. Note that there is a misprint
in the eq. (3.31) of the paper \cite{jmp95c}.


\section{Powers of Schr\"odinger-like operators}

The matrix elements of powers of a Schr\"odinger-like operator are already given 
in compact form by (\ref{50hhh}). 
Let us present here a more explicit expression for the powers of the proper Schr\"odinger 
operator (\ref{39c}).

\begin{theorem}
The matrix elements of the $k$-th power of the $r$-th order
Schr\"odinger-like operator (\ref{51c}) read
\be
<m|L_r^k|n>=<m+rk|n>
+\sum\limits_{d=1}^k\sum\limits_{|{\bf m}|=m-n+r(k-d)}
h_{r,k}({\bf m})u_{m_d}\cdots u_{m_1}
\label{106hhh}
\ee
where 
the inner sum here runs over non-negative integers $m_1\ge 0, \dots, m_d\ge 0$,
such that
\be
|{\bf m}|\equiv m_1+\cdots+m_d=m-n+r(k-d)
\ee
and
the numerical coefficients read
\be
h_{r,k}({\bf m})\equiv h_{r,k}(m_1,m_2,\dots,m_d)
=\sum_{i_1,\cdots,i_{d};}\prod_{p=1}^d
{{|{\bf m}|_p+n-r(i_{p}-p+1)\choose m_p}}.
\label{15f}
\ee
The sum in (\ref{15f}) runs over non-negative integers $i_1,\dots, i_d$
such that
\be
0\le i_1<i_2<\cdots<i_{d-1}<i_d<k,
\ee
\be
|{\bf m}|_p+n-r(i_{p}-p+1)\ge 0.
\ee

\end{theorem}

The proof is performed by the method of matrix elements, analogous to 
the proof of the Theorem 1 and, hence, is omitted.

Let us translate the result into the conventional notation.

\begin{corollary}
The $k$-th power of the $r$-th order
Schr\"odinger-like operator (\ref{51c}) applied to a function $f$
gives
\be
L_r^k f=f_{rk}
+\sum\limits_{d=1}^k\sum\limits_{|\widetilde{{\bf m}}|=r(k-d)}
h_{r,k}(\widetilde{{\bf m}})u_{m_d}\cdots u_{m_1}f_{m_0}
\label{107hhh}
\ee
where $f_m:={d^m\over dx^m}f$,
$\widetilde{{\bf m}}\equiv (m_0,{\bf m})\equiv (m_0,m_1,\dots,m_d)$,
the inner sum here runs over non-negative integers 
$m_0\ge 0, m_1\ge 0, \dots, m_d\ge 0$,
such that
\be
|\widetilde{{\bf m}}|\equiv m_0+m_1+\cdots+m_d=r(k-d)
\ee
and
the numerical coefficients read
\be
h_{r,k}(\widetilde{{\bf m}})\equiv h_{r,k}(m_0,m_1,m_2,\dots,m_d)
=\sum_{i_1,\cdots,i_{d};}\prod_{p=1}^d
{{|\widetilde{{\bf m}}|_p-r(i_{p}-p+1)\choose m_p}}.
\label{15fhh}
\ee
The sum in (\ref{15fhh}) is the same as in the theorem 4.

\end{corollary}

\paragraph{Proof.} There holds, with a mixture of matrix element notation and conventional
notation,
\be
L_r^kf=\sum_{0\le n\le rk}<0|L^k_r|n>f_n.
\ee
Here we insert $<0|L^k_r|n>$ from the theorem 4, i.e. we set 
$m=0$ there, rename $m_0:=n$ and define a new multiindex
$\widetilde{{\bf m}}=(m_0,{\bf m})$.

The numerical coefficients in (\ref{107hhh}) for $d=1$ and for $d=k$ exhibit a simple structure,
namely,
\be
h_{r,k}(m_0,m_1)=\sum_{i=0}^{k-1}{ri\choose m_1},
\ee
\be
h_{r,k}(m_0,m_1,\dots,m_k)=h_{r,k}(0,\dots,0)=1.
\ee
Thus
\be
L_r^kf=f_{rk}
+\sum_{m_0+m_1=r(k-1)}\sum_{i=0}^{k-1}{ri\choose m_1}u_{m_1}f_{m_0}
+\cdots+u^kf
\ee
where the dots indicate the terms of degree in $u$ greater than $1$ and less that $k$.

More explicit formulas for $L_r^kf$ are available for $r=1$. The powers of the first order
operator
\be
L_1={d\over dx}+u
\ee
are given by \cite{schimming96}
\be
L_1^kf=\sum_{l=0}^k{k\choose l}{\cal B}_l(u_0,\dots,u_{l-1})f_{k-l},
\ee
where ${\cal B}_l={\cal B}_l(y_1,\dots,y_l)$ $(l=0,1,2,\dots)$ denotes the {\it Bell
polynomials}, which we already discussed above.


\section*{Acknowledgements}

This work was supported by the Deutsche Forschungsgemeinschaft. 




\end{document}